# Priority Based Dynamic Round Robin (PBDRR) Algorithm with Intelligent Time Slice for Soft Real Time Systems

[#1] Prof. Rakesh Mohanty
[#2] Prof. H. S. Behera
Department of Computer Science & Engineering
Veer Surendra Sai University of Technology, Burla
Sambalpur, Orissa, India
[#1] rakesh.iitmphd@gmail.com
[#2] hsbehera_india@yahoo.com

[#3] Khusbu Patwari
[#4] Monisha Dash
[#5] M. Lakshmi Prasanna
Department of Computer Science & Engineering
Veer Surendra Sai University of Technology, Burla
Sambalpur, Orissa, India

*Abstract*—In this paper, a new variant of Round Robin (RR) algorithm is proposed which is suitable for soft real time systems. RR algorithm performs optimally in timeshared systems, but it is not suitable for soft real time systems. Because it gives more number of context switches, larger waiting time and larger response time. We have proposed a novel algorithm, known as Priority Based Dynamic Round Robin Algorithm(PBDRR), which calculates intelligent time slice for individual processes and changes after every round of execution. The proposed scheduling algorithm is developed by taking dynamic time quantum concept into account. Our experimental results show that our proposed algorithm performs better than algorithm in [8] in terms of reducing the number of context switches, average waiting time and average turnaround time.

I. INTRODUCTION

Real Time Systems (RTS) are the ones that are designed to provide results within a specific time-frame. It must have well defined fixed and response time constraints and the processing must be done within the defined constraints or the system will fail. RTS are basically divided into three types: hard, firm and soft. In hard real time systems, failure to meet deadline or response time constraints leads to system failure. In firm real time systems, failure to meet deadline can be tolerated. In soft real time systems, failure to meet deadline doesn't lead to system failure, but only performance is degraded[6]. Space research, weather forecast, seismic detection, audio conferencing, video conferencing, money withdrawal from ATM, railway and flight reservation etc are some of the applications of real time systems. The simple RR algorithm cannot be applied in soft real time systems as it gives longer waiting and response time. Yashuwaanth and et. al. [8] have proposed a scheduling algorithm for soft real time systems where Intelligent Time Slice(ITS) for all the processes has been calculated. The processes are scheduled using RR with ITS as time quantum. By taking dynamic time concept with ITS, we have proposed a new algorithm which gives improved performance than the algorithm proposed in [8].

A. *Real Time Scheduling Algorithms*

Some of the well known real-time scheduling algorithms are described as follows. **Rate Monotonic Algorithm(RM)** is a fixed priority scheduling algorithm which consists of assigning the highest priority to the highest frequency tasks in the system, and lowest priority to the lowest frequency tasks. At any time, the scheduler chooses to execute the task with the highest priority. By specifying the period and computational time required by the task, the behavior of the system can be categorized *apriori*. **Earliest-Deadline-First Algorithm (EDF)** uses the deadline of a task as its priority. The task with the earliest deadline has the highest priority, while the task with the latest deadline has the lowest priority. **Minimum-Laxity-First Algorithm (MLF)** assigns a *laxity* to each task in a system, and then selects the task with the minimum laxity to execute next. Laxity is defined as the difference between deadline by which the task must be completed and the amount of computation remaining to be performed. **Maximum-Urgency-First Algorithm (MUF)** is a combination of fixed and dynamic priority scheduling. In this algorithm each task is given an urgency which is defined as a combination of two fixed priorities, and a dynamic priority. One of the fixed priorities, called the criticality, has highest priority among the three, and then comes the dynamic priority which has precedence over the user priority (fixed priority). The dynamic priority is inversely proportional to the laxity of a task.

B. *Related Work*

In real time systems, the rate monotonic algorithm is the optimal fixed priority scheduling algorithm where as the earliest-deadline-first and minimum-laxity-first algorithms are the optimal dynamic priorities scheduling algorithms as presented by Liu and Layland in their paper [1]. S. Baskiyar and N. Meghanathan have presented a survey on contemporary Real Time Operating System (RTOS) which includes parameters necessary for designing a RTOS, its desirable features and basic requirements[6]. A dynamically reconfigurable system can change in time without the need to halt the system. David B. Stewart and Pradeep K. Khosla proposed the maximum-urgency-first algorithm, which can be used to predictably schedule dynamically changing systems [2]. The scheduling mechanism of the maximum-urgency-first may cause a critical task to fail. The modified maximum urgency first scheduling algorithm by Vahid Salmani, Saman Taghavi Zargar, and Mahmoud Naghibzadeh resolves the above mentioned problem [7]. C.Yashuwaanth proposed a Modified RR(MRR) algorithm which overcomes the limitations of simple RR and is suitable for the soft real time systems [8].

*C. Our Contribution*

In our work, we have proposed an improved algorithm as compared to the algorithm defined in [8]. Instead of taking static time quantum, we have taken dynamic time quantum which changes with every round of execution. Our experimental results show that PBDRR performs better than algorithm MRR in [8] in terms of reducing the number of context switches, average waiting time and average turnaround time.

*D. Organization of Paper*

Section II presents the pseudo code and illustration of our proposed PBDRR algorithm. In section III, Experimental results of the PBDRR algorithm and its comparison with the MRR algorithm is presented. Section IV contains the conclusion.

## II. OUR PROPOSED ALGORITHM

The early the shorter processes are removed from the ready queue, the better the turnaround time and the waiting time. So in our algorithm, the shorter processes are given more time quantum so that they can finish their execution earlier. Here shorter processes are defined as the processes having less assumed CPU burst time than the previous process. Performance of RR algorithm solely depends upon the size of time quantum. If it is very small, it causes too many context switches. If it is very large, the algorithm degenerates to FCFS. So our algorithm solves this problem by taking dynamic intelligent time quantum where the time quantum is repeatedly adjusted according to the shortness component.

*A. Our Proposed Algorithm*

In our algorithm, Intelligent Time Slice(ITS) is calculated which allocates different time quantum to each process based on priority, shortest CPU burst time and context switch avoidance time. Let the original time slice (OTS) is the time slice to be given to any process if it deserves no special consideration. Priority component (PC) is assigned 0 or 1 depending upon the priority assigned by the user which is inversely proportional to the priority number. Processes having highest priority are assigned 1 and rest is assigned 0. For Shortness Component(SC) difference between the burst time of current process and its previous process is calculated. If the difference is less than 0, then SC is assigned 1, else 0. For calculation of Context Switch Component (CSC) first PC, SC and OTS is added and then their result is subtracted from the burst time. If this is less than OTS, it will be considered as Context Switch Component (CSC). Adding all the values like OTS, PC, SC and CSC, we will get intelligent time slice for individual process.

Let '$TQ_i$' is the time quantum in round i. The number of rounds i varies from 1 to n, where value of i increments by 1 after every round till ready queue is not equal to NULL.

1. **Calculate ITS for all the processes present in the ready queue.**
2. **While(ready queue!= NULL)**
   ```
   {
   For i=1 to n do
     {
       if ( i ==1)
       {
         TQ_i = { ½ ITS, if SC= 0
                  ITS,   otherwise
       }
       Else
       {
         TQ_i = { TQ_{i-1} + ½ TQ_{i-1},   if SC=0
                  2 * TQ_{i-1},            otherwise
       }
       If (remaining burst time -TQ_i ) <=2
          TQ_i = remaining burst time
     } End of For
   } End of while
   ```
3. **Average waiting time, average turnaround time and no. of context switches are calculated**

   End

Fig-1: Pseudo Code of Proposed PBDRR Algorithm

*C. Illustration*

Given the CPU burst sequence for five processes as 50 27 12 55 5 with user priority 1 2 1 3 4 respectively. Original time slice was taken as 4. The priority component (PC) were calculated which were found as 1 0 1 0 0. Then the shortness component (SC) were calculated and found to be 0 1 1 0 1. The intelligent time slice were computed as 5 5 6 4 5. In first round, the processes having SC as 1 were assigned time quantum same as intelligent time slice whereas the processes having SC as 0 were given the time quantum equal to the ceiling of the half of the intelligent time slice. So processes P1, P2, P3, P4, P5 were assigned time quantum as 3 5 6 2 5. In next round, the processes having SC as 1 were assigned double the time slice of its previous round whereas the processes with SC equals to 0 were given the time quantum equal to the sum of previous time quantum and ceiling of the half of the previous time quantum. Similarly time quantum is assigned to each process available in each round for execution.

### III. EXPERIMENTS AND RESULTS

#### A. Assumptions

The environment where all the experiments are performed is a single processor environment and all the processes are independent. Time slice is assumed to be not more than maximum burst time. All the parameters like burst time, number of processes, priority and the intelligent time slice of all the processes are known before submitting the processes to the processor. All processes are CPU bound and no processes are I/O bound.

#### B. Experimental Frame Work

Our experiment consists of several input and output parameters. The input parameters consist of burst time, time quantum, priority and the number of processes. The output parameters consist of average waiting time, average turnaround time and number of context switches.

#### C. Data set

We have performed three experiments for evaluating performance of our new proposed PBDRR algorithm and MRR algorithm. We have considered 3 cases of the data set as the processes with burst time in increasing, decreasing and random order respectively. The significance the performance metrics for our experiment is as follows. *Turnaround time(TAT)*: For the better performance of the algorithm, average turnaround time should be less. *Waiting time(WT)*: For the better performance of the algorithm, average waiting time should be less. *Number of Context Switches(CS)*: For the better performance of the algorithm, the number of context switches should be less.

#### D. Experiments Performed

To evaluate the performance of our proposed PBDRR algorithm and MRR algorithm, we have taken a set of five processes in three different cases. Here for simplicity, we have taken 5 processes. The algorithm works effectively even if it used with a very large number of processes. In each case, we have compared the experimental results of our proposed PBDRR algorithm with the MRR algorithm presented in [8].

**Case 1:** We assume five processes arriving at time = 0, with increasing burst time (P1 = 5, P2 = 12, P3 = 16, P4 = 21, p5= 23) and priority (p1=2, p2=3, p3=1, p4=4, p5=5).

TABLE-1 ( MRR – Case 1)

| Process id | Burst time | Priority | OTS | PC | SC | CSC | ITS |
|---|---|---|---|---|---|---|---|
| P1 | 5 | 2 | 4 | 0 | 0 | 1 | 5 |
| P2 | 12 | 3 | 4 | 0 | 0 | 0 | 4 |
| P3 | 16 | 1 | 4 | 1 | 0 | 0 | 5 |
| P4 | 21 | 4 | 4 | 0 | 0 | 0 | 4 |
| P5 | 23 | 5 | 4 | 0 | 0 | 0 | 4 |

TABLE-2 ( PBDRR- Case 1)

| Process id | SC | ITS | ROUNDS | | | | |
|---|---|---|---|---|---|---|---|
| | | | $1^{st}$ | $2^{nd}$ | $3^{rd}$ | $4^{th}$ | $5^{th}$ |
| P1 | 0 | 5 | 5 | 0 | 0 | 0 | 0 |
| P2 | 0 | 4 | 2 | 3 | 7 | 0 | 0 |
| P3 | 0 | 5 | 3 | 5 | 8 | 0 | 0 |
| P4 | 0 | 4 | 2 | 3 | 5 | 8 | 3 |
| P5 | 0 | 4 | 2 | 3 | 5 | 8 | 5 |

TABLE – 3 ( Comparison between MRR and PBDRR)

| Algorithm | Average TAT | Average WT | CS |
|---|---|---|---|
| MRR | 51.2 | 35.8 | 19 |
| PBDRR | 46.4 | 31 | 17 |

The TABLE-1 and TABLE-2 show the output using algorithm MRR and our new proposed PBDRR algorithm. Table-3 shows the comparison between the two algorithms. Figure-2 and Figure-3 show Gantt chart for algorithms MRR and PBDRR respectively.

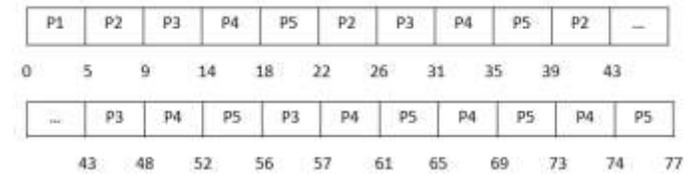

Fig. 2 : Gantt Chart for MRR(Case-1)

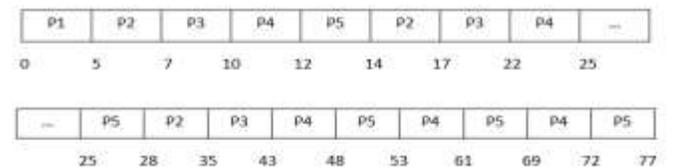

Fig. 3: Gantt Chart for PBDRR (Case-1)

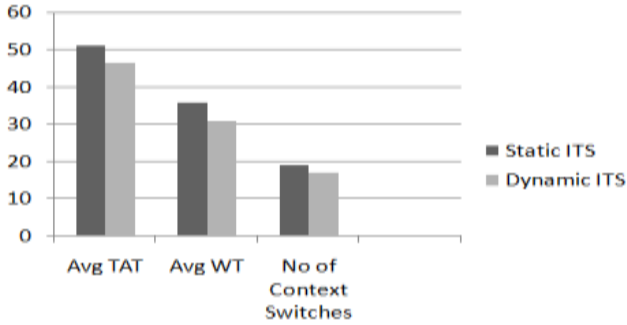

Fig.4 : Comparison of Performance of Algorithms - MRR with Static ITS and PBDRR with dynamic ITS ( Case-1 )

**Case 2:** We Assume five processes arriving at time = 0, with decreasing burst time (P1 = 31, P2 = 23, P3 = 16, P4 = 9, p5= 1) and priority (p1=2, p2=1, p3=4, p4=5, p5=3). The TABLE-4 and TABLE-5 show the output using algorithms MRR and PBDRR respectively. TABLE-6 shows the comparison between the two algorithms.

TABLE-4 ( MRR- Case 2)

| Process id | Burst time | Priority | OTS | PC | SC | CSC | ITS |
|---|---|---|---|---|---|---|---|
| P1 | 31 | 2 | 4 | 0 | 0 | 0 | 4 |
| P2 | 23 | 1 | 4 | 1 | 1 | 0 | 6 |
| P3 | 16 | 4 | 4 | 0 | 1 | 0 | 5 |
| P4 | 9 | 5 | 4 | 0 | 1 | 0 | 5 |
| P5 | 1 | 3 | 4 | 0 | 1 | 0 | 1 |

TABLE-5 ( PBDRR- Case 2)

| Process ID | Burst Time | SC | ITS | Rounds | | | |
|---|---|---|---|---|---|---|---|
| | | | | 1st | 2nd | 3rd | 4th |
| P1 | 31 | 0 | 4 | 2 | 3 | 5 | 21 |
| P2 | 23 | 1 | 6 | 6 | 12 | 5 | 0 |
| P3 | 16 | 1 | 5 | 5 | 11 | 0 | 0 |
| P4 | 9 | 1 | 5 | 5 | 4 | 0 | 0 |
| P5 | 1 | 1 | 1 | 1 | 0 | 0 | 0 |

TABLE – 6 ( Comparison between MRR and PBDRR)

| Algorithm | Avg TAT | Avg WT | CS |
|---|---|---|---|
| MRR | 54 | 38 | 18 |
| PBDRR | 50.4 | 34.4 | 12 |

Figure-5 and Figure-6 show Gantt chart for the algorithms MRR and PBDRR respectively.

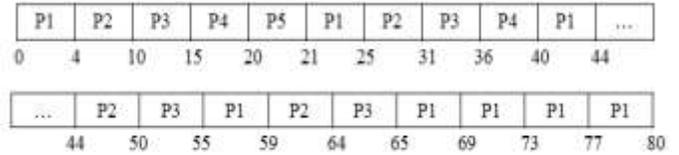

Fig. 5 : Gantt Chart for MRR(Case-2)

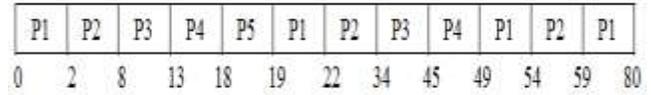

Fig. 6: Gantt Chart for PBDRR (Case-2)

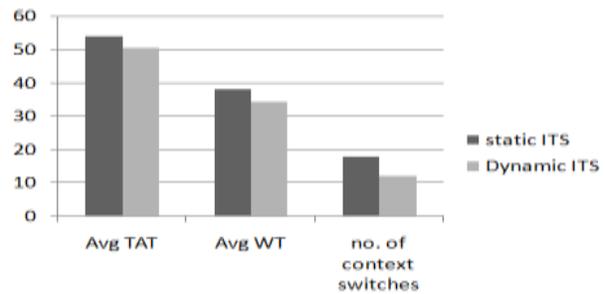

Fig. 7 : Comparison of Performance of Algorithms - MRR with Static ITS and PBDRR with dynamic ITS ( Case-2 )

**Case 3:** We assume five processes arriving at time = 0, with random burst time (P1 = 11, P2 = 53, P3 = 8, P4 = 41, p5= 20) and priority (p1=3, p2=1, p3=2, p4=4, p5=5). The TABLE-7 and TABLE-8 show the output using algorithms MRR and PBDRR respectively. TABLE-9 shows the comparison between the two algorithms. Figure-8 and Figure-9 show Gantt chart for both the algorithms.

TABLE-7 ( MRR- Case 3)

| Process id | Burst time | Priority | OTS | PC | SC | CSC | ITS |
|---|---|---|---|---|---|---|---|
| P1 | 11 | 3 | 4 | 0 | 0 | 0 | 4 |
| P2 | 53 | 1 | 4 | 1 | 0 | 0 | 5 |
| P3 | 8 | 2 | 4 | 0 | 1 | 3 | 8 |
| P4 | 41 | 4 | 4 | 0 | 0 | 0 | 4 |
| P5 | 20 | 5 | 4 | 0 | 1 | 0 | 5 |

TABLE-8 ( PBDRR- Case 3)

| Process id | SC | ITS | ROUNDS | | | | | |
|---|---|---|---|---|---|---|---|---|
| | | | 1st | 2nd | 3rd | 4th | 5th | 6th |
| P1 | 0 | 4 | 2 | 3 | 6 | 0 | 0 | 0 |
| P2 | 0 | 5 | 3 | 5 | 8 | 12 | 18 | 7 |
| P3 | 1 | 8 | 8 | 0 | 0 | 0 | 0 | 0 |
| P4 | 0 | 4 | 2 | 3 | 5 | 8 | 12 | 11 |
| P5 | 1 | 5 | 5 | 10 | 5 | 0 | 0 | 0 |

| Algorithm | Avg TAT | Avg WT | CS |
|---|---|---|---|
| MRR | 80.8 | 54.2 | 29 |
| PBDRR | 76 | 49.4 | 18 |

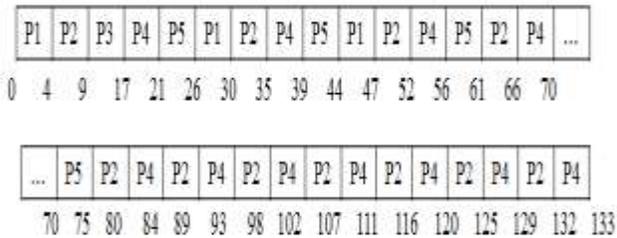

Fig. 8 : Gantt Chart for MRR(Case-3)

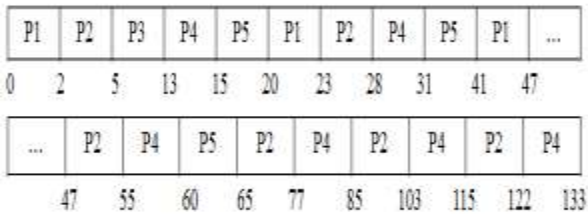

Fig. 9 : Gantt Chart for PBDRR (Case-3)

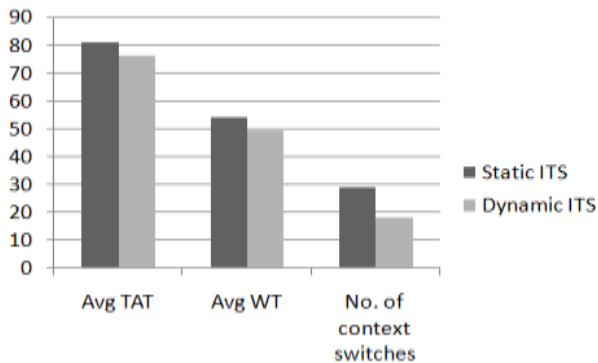

Fig.10 : Comparison of Performance of Algorithms - MRR with Static ITS and PBDRR with dynamic ITS ( Case-3 )

## IV. CONCLUSION

From the above comparisons, we observed that our new proposed algorithm PBDRR is performing better than the algorithm MRR proposed in paper [8] in terms of average waiting time, average turnaround time and number of context switches thereby reducing the overhead and saving of memory spaces. In the future work, deadline can be considered as one of the input parameter in addition to the priority in the proposed algorithm. Hard Real Time Systems have hard deadline, failing which causes catastrophic events. In future work, a new algorithm in hard real time systems with deadline can be developed.

## REFERENCES

[1] C. L. Liu and James W. Layland : *Scheduling Algorithms for Multiprogramming in a Hard-Real-Time Environment,* Journal of the ACM(JACM), Vol. 20, Issue 1, January, 1973.

[2] David B. Stewart and Pradeep K. Khosla: *Real-Time Scheduling of Dynamically Reconfigurable Systems,* Proceedings of the IEEE International Conference on Systems Engineering, pp 139-142, August, 1991.

[3] Krithi Ramamrithm and John A. Stankovic*: Scheduling Algorithms and Operating System Support for Real Time Systems*, Proceedings of the IEEE, Vol. 82, Issue 1, pp 55-67, January- 1994.

[4] R. I. Davis and A. Burns : *Hierarchical Fixed Priority Pre-emptive Scheduling,* Proceedings of the 26th IEEE International Real-Time Systems Symposium(RTSS), pp 389-398, 2005.

[5] Omar U. Pereira Zapata, Pedro Mej´ıa Alvarez: *EDF and RM Multiprocessor Scheduling Algorithms: Survey and Performance Evaluation*, Technical Report, 1994.

[6] S. Baskiyar and N. Meghanathan: *A Survey On Contemporary Real Time Operating Systems,* Informatica, 29, pp 233-240, 2005.

[7] Vahid Salmani, Saman Taghavi Zargar, and Mahmoud Naghibzadeh*: A Modified Maximum Urgency First Scheduling Algorithm for Real-Time Tasks*, World Academy of Science, Engineering and Technology. Vol. 9, Issue 4, pp 19-23, 2005.

[8] C. Yaashuwanth and R. Ramesh, : *A New Scheduling Algorithm  for Real Time System,* International Journal of Computer and Electrical Engineering (IJCEE), Vol. 2, No. 6, pp 1104-1106, December, 2010.

AUTHORS PROFILE

Prof. Rakesh Mohanty is a Lecturer in Department of Computer Science and Engineering, Veer Surendra Sai University of Technology, Burla, Orissa, India. His research interests are in operating systems, algorithms and data structures .

Prof. H. S. Behera is a Senior Lecturer in Department of Computer Science and Engineering, Veer Surendra Sai University of Technology, Burla, Orissa, India. His research interests are in operating systems and data mining .

Khusbu Patwari, Monisha Dash and M. Lakshmi Prasanna have completed their B. Tech. in Department of Computer Science and Engineering, Veer Surendra Sai University of Technology, Burla, Orissa, India in 2010.